\documentclass[12pt,preprint]{aastex} 

\usepackage{dcolumn}
\usepackage{bm}

\begin{document}


\title{\bf Long-term Radio Modulation in Sagittarius A* from Spin-Induced Disk Precession}

\author{Martin Prescher$^{1,2,3}$ and  Fulvio Melia$^{1}$}
\affil{$^1$Physics Department and Steward Observatory,\\
The University of Arizona, Tucson AZ 85721}
\affil{$^2$Fachbereich Physik,\\Freie Universit\"at Berlin, Arnimallee 14, 14195 Berlin}
\affil{$^3$Fulbright Fellow}

\begin{abstract}
There is some evidence, though yet unconfirmed, that Sagittarius A*---the supermassive
black hole at the Galactic center---emits its radio waves modulated with a $\sim 100$-day
period. What is intriguing about this apparent quasi-periodicity is that, though
the amplitude of the modulation increases with decreasing wavelength (from $3.6$ to
$1.3$ cm), the quasi-period itself does not seem to depend on the frequency of the
radiation. It is difficult to imagine how a binary companion, were that the cause
of this modulation, could have escaped detection until now. Instead, it has been
suggested that the spin-induced precession of a disk surrounding a slowly rotating
black hole could have the right period to account for this behavior.  In this paper,
we examine how Sagittarius A*'s light curve could be modulated by this mechanism.
We demonstrate that the partial occultation of a nonthermal halo by a compact,
radio-opaque disk does indeed produce the observed frequency-dependent amplitude.
This appears to be in line with other observational arguments suggesting that
Sagittarius A*'s mm/sub-mm spectrum is produced by a $\sim 10$ Schwarzschild-radius
disk, whereas its cm-waves originate from a nonthermal particle distribution in
a halo extending out to over $20$ Schwarzschild radii. Interestingly, this model
suggests that the observed period corresponds to half the precession period and
that a non-axisymmetric disk could produce a second period roughly twice as long
as the first.
\end{abstract}

\keywords{accretion, accretion disks --- black hole physics --- Galaxy: center
 --- gravitation --- radiation mechanisms: nonthermal --- relativity}

\maketitle


\section{Introduction}
The weekly VLA monitoring of Sagittarius A*, the $\sim 4\times 10^6\;M_\odot$ supermassive
black hole at the center of our Galaxy, has accumulated over 20 years of variability data at
$1.3$, $2.0$, $3.6$, $6.0$, and $20$ cm wavelengths.  The sampling within this period has
been somewhat irregular (Zhao, Bower, and Goss 2001).  Nonetheless, the power spectral
density (PSD) reveals a clear peak near $1\times 10^{-7}$ Hz, with a progressively smaller
significance at longer wavelengths.  This frequency corresponds to a periodic
modulation of approximately $100-120$ days; the actual best-fit period extracted
from the combined data sets is $106$ days.

This result is at once intriguing and unsettling. The fact that a $\sim 17$-minute Keplerian
period has been seen in Sagittarius A*'s infrared emission (Genzel et al. 2003) makes this
cyclic modulation easier to accept.  Yet the implied radio period ($\sim 106$ days)
contrasts sharply with the dynamical time scale ($\sim 17-20$ minutes) associated
with motion in the inner disk (see Melia and Falcke 2001 for a recent review).

Perhaps the periodic radio signal is a false detection due to a combination
of a random process and the irregular sampling pattern.  However Monte Carlo
tests with data created from various sources of noise using this same sampling
don't seem to bear this out.  Regardless of the type of noise used in the
simulations---including white noise, Gaussian noise around a mean, and a Poisson
distribution of flares---the probability of false detection due to any such
random process appears to be less than $5\%$. (However, one should keep in mind that
$1/f$---or red---noise could in principle mimic such a periodic signal if the random
fluctuations have an appropriate {\it scale}; G. Bower, private communication.)
Perhaps also supporting the view that this period may be real is the observation
that both the absolute ($\Delta S$) and fractional ($\Delta S/S)$ amplitudes of the
pulsed component increase toward shorter wavelengths, yet the period appears to be
independent of wavelength.

The $106$-day cycle evident primarily at $1.3$ and $2.0$ cm may be a valuable
tool for probing Sagittarius A*'s inner workings should it truly have something
to do with the source.  High-resolution VLA observations have already
ruled out the possibility that such a period might be produced by an
orbiting emitting object (Bower and Backer 1998).  The $106$-day
orbit of a companion to Sagittarius A* would have a radius $\approx 60$
A.U., corresponding to an angular separation of $\approx 8$ milliarcseconds at
$8$ kpc.  A compact $0.2$-Jy source separated from Sagittarius A* by this amount
would have easily been detected with the VLBA at wavelengths shorter than
$3.6$ cm.  The unlikelihood of Sagittarius A* having an orbiting
companion is further supported by its observed lack of proper motion,
which precludes any possible association with rapidly moving components.  In addition,
a stellar origin for such a source would fall well short of the power required to
account for the measured radio emission. All in all, the evidence seems to favor an
interpretation in which the $106$-day periodic variations, if real, are intrinsic to
Sagittarius A* itself.

The characteristics of this 106-day cycle constrain the nature of its origin rather
tightly. First, the observed period is, as we have said, independent of wavelength.
The emission in Sagittarius A* at different frequencies is produced on
different spatial scales (see, e.g., Melia, Jokipii, and Narayanan 1992), so
the period should be induced by a single process. Otherwise, we would expect to
see different periods at different frequencies. What is required is
something that can cause correlated fluctuations across a broad range of wavelengths.
Second, the period is four orders of magnitude longer than the dynamical time scale in
the inner disk surrounding Sagittarius A*. Could it be produced on a much larger
spatial scale? Sagittarius A*'s 2-cm emission is produced within a region no bigger
than $\sim 24$ A.U. (Krichbaum et al. 1999), for which the corresponding dynamical
time scale is calculated to be about one and a half days; so the answer is apparently
no. Higher frequency emission is produced within still smaller regions, associated with
even smaller time scales.

We may ask then, whether this modulation could be produced by a corrugation wave in
an accretion disk, which is used to account for the quasi-periodic oscillations (QPOs)
seen in low-mass X-ray binaries. These waves have periods that are much longer than
the corresponding dynamical time scale (Kato 1990), but they depend on
radius and thus may not be able to account for the first feature described above.
Moreover, Sagittarius A*'s light curves show quite stable periodic fluctuations,
rather than the uncorrelated segments constituting QPOs.  The evidence is pointing
to a single process evolving in a relatively confined region, certainly no bigger than
100 Schwarzschild radii ($r_S\equiv 2GM/c^2\approx 10^{12}$ cm for a $4\times 10^6\;
M_\odot$ black hole).

In an earlier paper (Liu and Melia 2002), we noted that the gravitational acceleration
in a Kerr metric acquires a dependence on poloidal angle (relative to
the black hole's spin axis), so that matter orbiting above or below the equatorial
plane experiences a restoring force toward the equator. This results in the precession
of its angular momentum vector about the black hole's axis of rotation.  The physical
conditions in Sagittarius A*'s Keplerian region lead to strong coupling between
neighboring rings, and the $\sim 20-30\;r_S$ compact disk therefore torques more or less
as a rigid body. Under an appropriate set of circumstances (see Liu and Melia 2002),
the precession period can exceed a hundred days, and the long-term radio modulation in
Sagittarius A* may be closely related to its short-term X-ray and infrared variability
after all---via the dynamical properties of the disk in a rotating spacetime.

However, whereas this earlier treatment established the viability of a spin-induced
disk precession in accounting for the 106-day modulation, it left open the question
of how exactly this periodicity is manifested.  The purpose of this paper is to
demonstrate that the observed fluctuations, frequency-dependent amplitudes, and
periodicity can in fact be produced as a result of partial occultation of a nonthermal
halo surrounding Sagittarius A* by the pivoting disk. As we shall see, the properties
inferred for this source based on its spectrum and polarization characteristics (see
Liu and Melia 2001), produce a surprisingly accurate fit to the observed radio lightcurve,
its modulated amplitude, and the frequency-dependent signal strength. In \S\ 2 we describe
the method used for this analysis, including the source geometry and particle properties.
We summarize our results in \S\ 3, and present our conclusions in \S\ 4.


\section{The Method}
The level of polarization seen in Sagittarius A* at mm/sub-mm wavelengths approaches
$7-10\%$ (Aitken et al. 2000). However, this object reveals a lack ($<1\%$) of
linear polarization below $112$ GHz, though some circular polarization
($\sim 1\%$) has been detected (Bower et al. 1999; Bower et al. 2001).
These prominent spectral and polarimetric differences (Melia, Bromley,
\& Liu 2001) between the cm and the mm/sub-mm bands suggest two different
emission components in Sgr A*. As we have already noted, higher frequencies
correspond to smaller spatial scales (see also Narayan et al. 1995), so the
mm/sub-mm radiation is likely produced in the vicinity of the black hole.
Earlier work (e.g., Melia 1992, 1994; Coker \& Melia 1997) has indicated that Sgr A*
is accreting from the stellar winds surrounding the black hole and that the
infalling gas circularizes at a radius of $\sim 20-800\ r_S$. Recent work on
Sagittarius A*'s emissivity (Melia, Liu, \& Coker 2001; Bromley et al. 2001)
has demonstrated that the inner $10\,r_S$ of the resultant Keplerian structure
can not only account for the mm/sub-mm properties via thermal synchrotron
emission, but it may also produce Sgr A*'s X-ray spectrum in the quiescent
state (Baganoff et al. 2001) via Comptonization of the mm/sub-mm photons.
On the other hand, the cm radio emission appears to be produced by non-thermal
synchrotron emission (Liu \& Melia 2001).

The integrated $\sim 1-20$ cm luminosity of Sgr A* is comparable to
the power extracted from its spin energy via a Blandford-Znajek type of
electromagnetic process if $a/M\approx 0.1$ (Liu and Melia 2002).  The 106-day
modulation could therefore presumably arise when the precessing disk periodically
shadows the non-thermal particles flooding the region surrounding the black hole
as they escape from their creation site near the event horizon, essentially forming
an expanding halo of relativistic particles. Some observational evidence for this
has recently been provided by VLBA closure amplitude imaging techniques at 7 mm
(Bower et al. 2004), which point to an intrinsic source size of $\approx 25\,r_S$
for Sagittarius A* at this wavelength.

We will therefore adopt the basic model displayed in Figure 1.
The compact disk (with outer radius $R_d$) that produces the mm/sub-mm emission
is opaque to radiation longward of $\sim 1$ cm, which we assume originates from
the surrounding, semi-transparent halo with radius $R_h$. To accurately determine the disk shadowing
effect, we integrate the non-thermal synchrotron emissivity $j_\nu$ (erg cm$^{-3}$
s$^{-1}$ ster$^{-1}$ Hz$^{-1}$) along each given line-of-sight (see Figure 1),
including the effects of opacity, such that
\begin{equation}\label{inu}
   I_\nu(s) = \int_{s_0}^s j_\nu(\sigma) \> e^{-\tau_\nu(\sigma)} \> \hbox{d}\sigma\;.
\end{equation}
In this expression,
\begin{equation}\label{depth}
   \tau_\nu(s) = \int_{s_0}^s \alpha_\nu(\sigma) \> \hbox{d}\sigma
\end{equation}
is the optical depth in terms of the absorption coefficient $\alpha_\nu(\sigma)$ $($cm$^{-1})$.
The flux is then calculated by integrating $I_\nu$ over all solid angles pertaining
to the source:
\begin{equation}\label{flux}
    F_\nu = \int I_\nu \> \cos \theta \> \hbox{d}\Omega,
\end{equation}
where $\theta$ is the angle between the line of sight to the black hole (at the center of
this geometric configuration) and the emitting area associated with $d\Omega$.

Numerically, we calculate the flux on a grid (see Figure 2), assuming $F_\nu$ to be
constant over each mesh element, so that the total flux may be written
\begin{equation}\label{flux2}
   F_\nu \approx \frac{\epsilon^2}{d^2} \sum_{i,j} I^{ij}_\nu\;,
\end{equation}
where $d$ is the distance to the Galactic center, $\epsilon$ is the side length
of each mesh element, and $\theta\approx 0$ for $\epsilon/d \ll 1$.

\subsection{Power Law Distribution}

We assume that within the halo the emitting particles form a power-law
distribution
\begin{equation}
N(E) \> \hbox{d}E = N_0 \> E^{-p} \> \hbox{d}E\;,
\end{equation}
with $p = 2$.  Thus,
\begin{equation}\label{j}
   j_\nu = \frac{\sqrt{3}}{4\pi} \frac{e^3}{mc^2} \> B \> \sin \psi \>
\int_{0}^\infty \!\! N(E) \> F(x) \> \hbox{d}E\;,
\end{equation}
where $B$ is the magnetic field strength, $\psi$ is the pitch
angle (the angle between the electron's velocity and $\mathbf{B}$),
$F(x) = x \int_x^\infty K_{5/3}(\xi) \> \hbox{d} \xi$, and $K_{5/3}$ is the
modified Bessel function of the second kind. Similarly,
\begin{equation}\label{alpha}
    \alpha_\nu = -\frac{\sqrt{3}e^3}{8 \pi m \> \nu^2} \> B \> \sin \psi \>
\int_0^\infty \!\! E^2 \frac{\partial}{\partial E} \left( \frac{N(E)}{E^2} \right)
F(x) \> \hbox{d} E.
\end{equation}

Equations (\ref{j}) and (\ref{alpha}) then become
\begin{eqnarray}\label{j_nu}
    j_\nu &=& C_1 \> N_0 \> \left( B \> \sin \psi \right)^{\frac{3}{2}}
\nu^{-\frac{1}{2}}\;, \\ \label{alpha_nu}
    \alpha_\nu &=& C_2 \> N_0 \> \left( B \> \sin \psi \right)^{2} \nu^{-3}\;,
\end{eqnarray}
where
\begin{eqnarray}\label{}
   C_1 &\equiv& \frac{13}{48 \pi^{3/2}} \>  \Gamma\left( \frac{5}{12} \right)
\Gamma\left( \frac{13}{12} \right) \> \sqrt{\frac{e^7}{2 m^5 c^9}}\;, \\
   C_2 &\equiv& \frac{e^4}{6 \pi \> m^4 c^5 } \;.
\end{eqnarray}
Equations (\ref{j_nu}) and (\ref{alpha_nu}) are used in Equations (\ref{inu}) through
(\ref{flux2}) to calculate the total flux.

\subsection{Parameters and model assumptions}

For the calculations described here, the disk precession is handled analytically,
with an orientation prescribed as a function of time, based on our previous work
(Liu and Melia 2002). A full SPH simulation is currently underway, and a more realistic
time-dependent geometric profile will be published elsewhere (Rockefeller,
Fryer, and Melia 2005). A key concern is whether the disk succumbs to the Bardeen-Pettersen
effect, in which the radially-dependent precession frequency can lead to dissipation between
neighboring rings, ultimately forcing the inner $\sim 10-20\,r_S$ of a disk inclined to the
black hole's spin axis to eventually settle into the equatorial plane.  It is commonly
thought that an accretion disk surrounding a spinning black hole must be warped,
with an overall angular momentum vector possibly misaligned relative to the spin axis, but
with its inner portion fully flattened at the equator.  But detailed hydrodynamical simulations
are now showing that this effect, although fully realized under a majority of physical
conditions, can be negated in cases where other couplings between neighboring rings in the
disk are strong. An example of this occurs when the disk plasma has a small Mach number
(typically $< 5$), for which large pressure gradients can then distribute the radially-dependent
precession torque and force the disk to rotate as a rigid body (see, e.g., Nelson \& Papaloizou
2000). In the case of Sgr A*, we estimate from the spectrum and polarization properties
of its disk that the Mach number in this system is $\sim 3$. Not surprisingly then, the
latest SPH simulations confirm the earlier semi-analytic results that the small disk in
Sgr A* is probably precessing as a rigid body.  In our calculation, we therefore evolve the
disk orientation analytically in time, assuming rigid body precession, and recalculate the
flux at $1.3$, $2.0$, and $3.6$ cm for each step.

\begin{figure}[ht]
\plotone{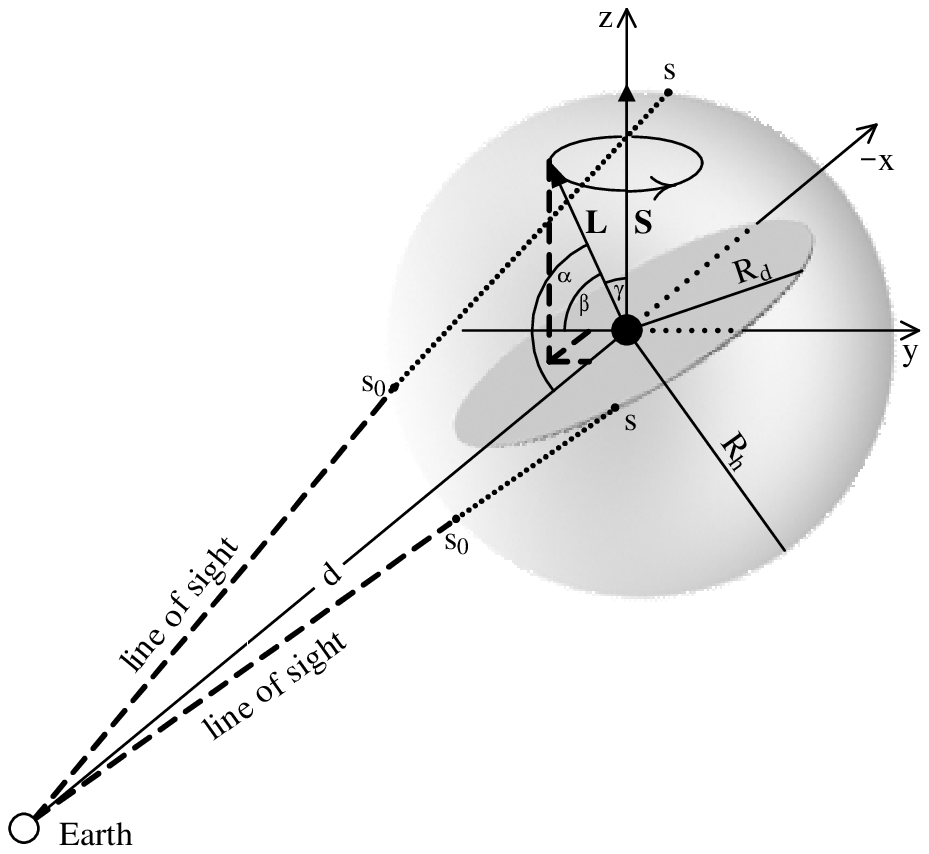}
\caption{The geometry used in this model. Here, $d$ denotes the distance to the
Galactic center, $R_h$ is the radius of the nonthermal halo surrounding Sgr A*, $R_d$ the
radius of the compact accretion disk, $\mathbf{S}$ is the spin vector of the black hole,
and $\mathbf{L}$ is the angular momentum vector of the disk making angles $\alpha$, $\beta$, and
$\gamma$ to the $x$, $y$, $z$ axes, respectively.}
\label{fig:1}
\end{figure}

As we can see from Equations (\ref{j_nu}) and (\ref{alpha_nu}), the model contains
five basic parameters: $N_0$, $B$, $R_h$, $R_d$ and $\gamma$. Here $\gamma$ is the angle
between the angular momentum vector of the disc and the spin axis of the black hole (see
Fig. \ref{fig:1}). The other two angles $\alpha$ and $\beta$ change in every time step.
The pitch angle $\psi$ depends on
distance $\sigma$ in Equation (\ref{inu}). For relativistic electron energies, we can
approximate $\psi \approx \vartheta$, where $\vartheta$ is the angle between the
magnetic field vector and the line of sight.  For the sake of simplicity, $N_0$ is
 taken to be a power-law function $\propto r^{-\varepsilon}$ (with $\varepsilon =
0,1,2$), in terms of $r$, the distance from the center of the halo. The magnetic field $B$
is assumed constant throughout the halo. Consequently,
there are effectively six overall parameters in this model:
$N_0(r=R_h)$, $B$, and $\varepsilon$, $R_h$, $R_d$ and $\gamma$.

Throughout this work the accretion disk is considered to be optically thick to
all radiation onward of $1$ cm, so any ray of light intercepting the disk is
immediately stopped at that point. Since the distance to the Galactic center $d$
is very large compared to the radius of the halo we may also assume parallel lines
of sight.

We have explored models with $R_d\sim17\,r_S$, $R_h\sim 22\,r_S$ for the halo, and a tilting
angle $\gamma$ between $\mathbf{L}$ and $\mathbf{S}$ of $\sim 60^\circ$ (or $30^\circ$ measured
from the xy-plane).  Based on earlier spectral fitting calculations, we know that typical values
for $N_0(r=R_h)$ are of order $10^{14} \> \hbox{cm}^{-3}$ erg$^p$. For a given
power-law dependence of $N_0$, this then also fixes the density throughout the emitting region.
Also based on earlier spectral fitting calculations (see Liu and Melia 2001), we infer a
magnetic field intensity of order $1$ G.

We point out, however, that the specific choice of the function $N_0(r)$ has little effect on the
shape of the light curve (as we shall see in Figs. 3 to 8 below), though the calculated flux does change
in response to the changing column depth through the emitting medium. Consequently, we have
chosen the model assumptions to be as simple as possible. A uniform electron distribution with
$\varepsilon = 0$, though unrealistic, allows a first critical analysis of the shadowing effect.

\section{Results}

A good fit to the folded light curves (shown in Figs. 3 and 4) is possible with either a uniform
electron distribution, or a power law. For the sake of specificity, we here focus on the results
produced with a uniform halo (Figs. 3, 4, and 5), and one in which $N_0(r)\propto 1/r$ (Figs. 6, 7,
and 8). We have also found that the light curve produced as a result of shadowing (dark solid
curve in these figures) is only weakly dependent on the absolute value of $N_0(R_h)$ and $B$,
which instead affect primarily the absolute flux level.  The shadowing effect is therefore
relatively independent of the details of the underlying model; it appears to be a rather robust
phenomenon.

In Figure \ref{fig:3}, we show the calculated light curve in comparison with the data at $1.3$ cm.
This fit was produced with $N_0(R_h) = 3.6 \times 10^{13} \hbox{ cm}^{-3}$ erg$^2$,
$B = 0.8 \hbox{ G}$, $R_h = 21.5\, r_S$ and $R_d = 17.5\,r_S$. Every data point on the solid
line represents the flux (in Jy) obtained for a
certain position of the accretion disk, which precesses around the (fixed) spin vector of the black
hole. In every time step between time $0$ days (which also corresponds to $0^\circ$) and $106$
days (which corresponds to an angle of $180^\circ$ of $\mathbf{L}$ with respect to the starting
point) the precessing angle is increased. At the maximum ($\approx 84$ days), the accretion disk
is seen edge-on and consequently, since the disk is here assumed to have no thickness, the integration
over the synchrotron emissivity extends over the full extent of the halo. In order to illustrate this,
for a tilting angle of $\gamma = 90$ degrees, i.e., $\mathbf{L}$ parallel to the $x$-axis in Figure
\ref{fig:1}, the curve produced by our simulation would be constant at the value of the maximum in
Figure \ref{fig:3}, since the accretion disk would be seen edge-on in every time step.

It should be emphasized that in Figure \ref{fig:3}, the mean flux has not been adjusted vertically,
but rather is set by the parameter values suggested by earlier spectral studies. One of the principal results
of our simulation is that the parameters that yield this correct flux, together with a reasonable
tilting angle of $60^\circ$, yield both the correct amplitude of the observed modulation, and its
variation in time.

In Figure 3, we also show the corresponding light curve for $2.0$ cm, using the same parameters as those
for $1.3$ cm. Notice that here the shadowing is less effective, which is due to two factors: first,
the halo becomes progressively more opaque with increasing wavelength, which means that a greater
fraction of the observed intensity arises from the medium in front of the disk; second, the emitting
volume contributing to the overall flux increases, rendering the fraction of the halo occulted by
the disk less significant with increasing wavelength. The optical depth through the halo at $1.3$ cm
is $\tau_\nu \approx 10^{-1}-10^{-2}$, so most of the photons pass through the medium unaffected.
However, $\tau_\nu$ goes as $\nu^{-3}$ (see Eq. \ref{alpha_nu}); at $2.0$ cm, it is $> 10^{-1}$.
We note that both the shape of the light curve and the amplitude of modulation fit the data just
as well at $2.0$ cm as they do at $1.3$ cm, for the {\it same} system parameters.

By 3.6 cm, $\tau_\nu$ has increased to values $> 1$ and, not surprisingly, most of the modulation due
to the precessing disk, which lies below the photosphere at this wavelength, has disappeared
(see Fig. \ref{fig:5}). This trend is reflected in the data as well, though we note that our calculated
flux is a factor 2 lower than that observed.  We attribute this to the fact that for simplicity
we have assumed a fixed halo size.  In reality, the fact that $\tau_\nu$ is increasing with increasing
wavelength also means that the emitting volume contributing to the overall flux must itself
increase. It would be a simple matter to reproduce the correct flux at $3.6$ cm by making $R_h$
wavelength dependent, but this would violate our goal of keeping this model as simple as possible.  In principle,
a full numerical simulation of the disk precession and particle acceleration and escape should yield
the correct behavior of quantities such as $N_0$ with radius, which would then make this point moot.

Figure 4 demonstrates the same effects as those discussed in Figure 3, except now
for a halo with uniform magnetic field, but with a density $N_0(r)\propto 1/r$, with
$N_0(R_h)=2.45 \times 10^{13} \hbox{ cm}^{-3}$ erg$^2$ and with $B=0.8 \hbox{ G}$.  The fact that a reasonable
fit to the data may be made with such disparate halo geometries affirms the robustness of this model.
However, changing the parameter values does produce quantitative variations in the modulation amplitude,
and its dependence on wavelength. Figure \ref{fig:5} shows the sensitivity of our simulation to changes
in $N_0(R_h)$. We see that the shape of the curve (at $1.3$ cm) is not affected. This is hardly surprising,
since $\tau_\nu < 1$ for all these densities. However, since $j_\nu \propto N_0$, the overall flux does
change with increasing $N_0$.

The consequence of changing the magnetic field is shown in Figure \ref{fig:5}. As was the
case for $N_0$, the shape of the curve does not depend on this value. Also, since the dependence
of $\alpha_\nu$ and $j_\nu$ on $B$ is only to the power $3/2$, the impact on the average flux is
also less pronounced.

The impact of parameter variations on the light curve at other frequencies is similar to that
for $1.3$ cm, so we won't show these here.

Changes in the disk's outer radius produce the modified modulations shown in Figure \ref{fig:5}.
All the curves coincide at the peak since the edge-on disk does not shadow any part of the halo.
Also, the amplitude decreases as the disk becomes smaller.

\section{Conclusions}

The picture described here works rather well in accounting for (1) the absolute flux, (2) the
observed modulation, and (3) the amplitude of the observed peaks in the $1.3$ and $2.0$ cm light
curves of Sagittarius A*.  However, future work should include a more realistic time-dependent
disk profile, the halo, and a more detailed exploration of possible magnetic field configurations.
For example, in this paper, we have restricted our attention exclusively to isotropic halos surrounding
Sgr A*, though we have allowed for a possible radial variation in its physical conditions, such as
the particle number density and the magnetic field. Clearly, though, one might expect some impact
on the results should the halo not be spherical, but rather toroidal, or cylindrical.  But as we
have already indicated earlier, the halo is optically thin to radio emission, and the dominant effect
is the shadowing of this emitter by the disk.  Our brief survey of the various possible geometries
has indicated that the {\it shape} of the modulated light curve is only weakly dependent on the
halo's structure.  Nonetheless, the absolute flux level {\it is} indeed sensitive to the halo's
geometry.  Some of the additional work necessary to fully explore the range of possibilities
(e.g., with SPH simulations) is currently underway and will be reported elsewhere.

Finally, we point out here something that was overlooked in Liu and Melia (2002). This is
the fact that if the disk precession is responsible for the modulation, then the observed period should
actually correspond to half of the precession period because of symmetry. This has some impact on the
inferred value of the black hole spin $2a/r_S$.  In addition, it may be possible to see two periods,
the primary one associated with half the precession period, and a second at roughly twice the first,
corresponding to the full period. The latter would arise if the disk were not perfectly symmetric,
so that the tilt in the first half of a cycle is not exactly the same as the tilt in the second.
This is a prediction of the model presented in this paper and could be included in future work.
Interestingly, there may already be some evidence for a second period, roughly twice as long as the first, in
data acquired since 2000 (Zhao, private communication).

To close, it is still not entirely clear whether this $\sim 100$-day period in Sagittarius A*'s
emission is in fact due to an intrinsic modulation. However, we have now shown that a spin-induced
disk precession for a slowly rotating black hole can account for this long period and, in addition,
that the partial occultation of a $\sim 1-3.6$-cm emitting halo by the radio-opaque precessing
disk can also produce the correct amplitude and time-dependent light curves
seen at $1.3$ and $2.0$ cm.  The halo becomes optically thick longward of $\sim
2.0$ cm, so the shadowing by the disk can no longer produce a modulation at these wavelengths.
Future work observationally should focus more carefully on the question of whether red noise
fluctuations could mimic the effect of a long periodic modulation in this source. Theoretically,
future work will determine the temporal profile of the precessing disk more accurately, and will
provide us with a better grasp of the halo's internal structure.

\begin{acknowledgments}
M.P. would like to thank the German Fulbright Commission for support far beyond the financial aspect of
academics, which made this work possible. Furthermore he likes to thank Steward Observatory for kind
hospitality and Casey Meakin and Philipp Strack for helpful discussions. This work was supported by
NASA grant NAG5-9205 and NSF grant 0402502 at the University of Arizona.
\end{acknowledgments}

\newpage

\newpage

\begin{figure}[ht]
\epsscale{0.5}
\plotone{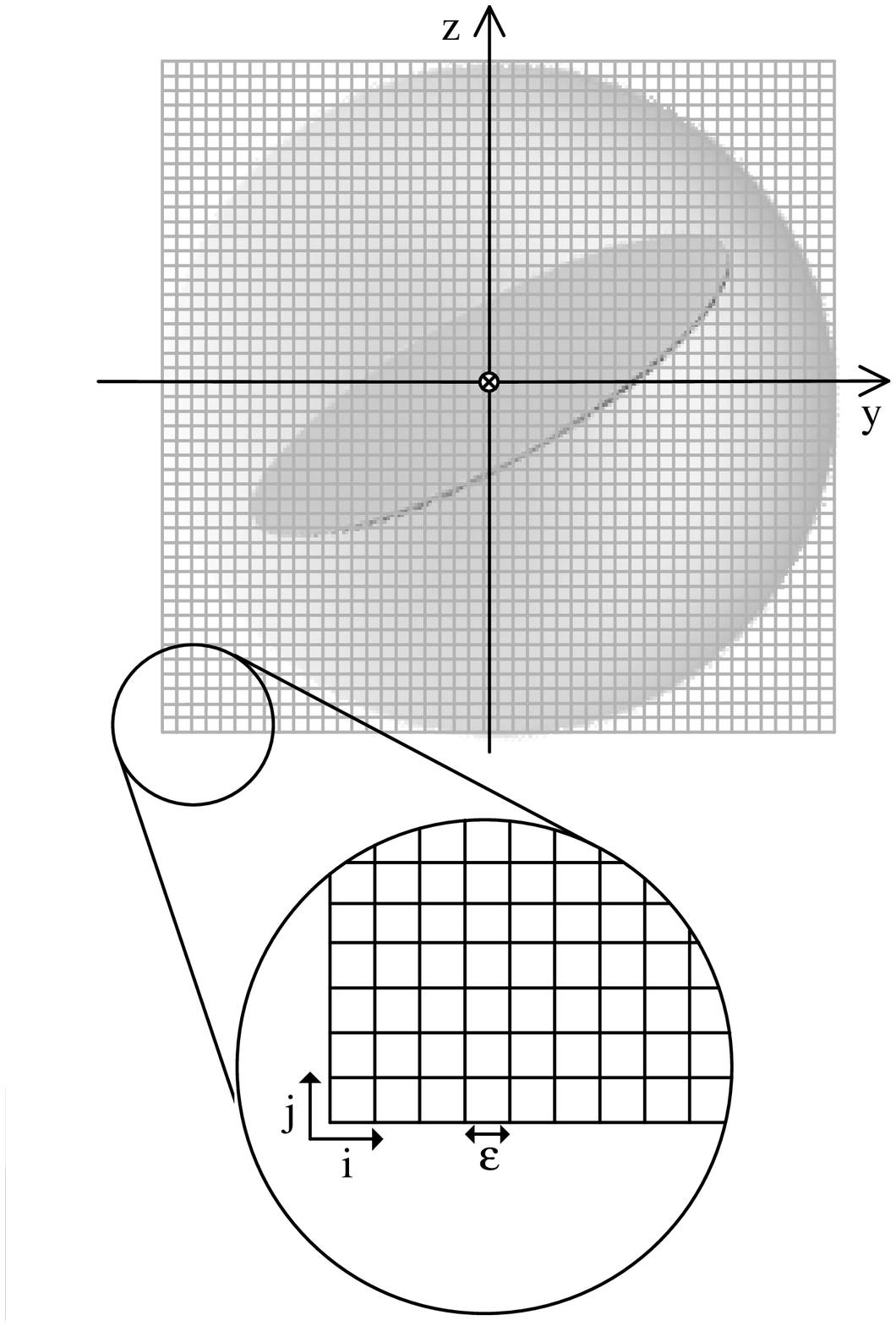}
\caption{The mesh used to calculated the intensity and flux measured at Earth.
In this paper, we have employed $100\times100$ grid points.}
\label{fig:2}
\end{figure}

\newpage
\begin{figure}[ht]
\epsscale{1.0}
\plotone{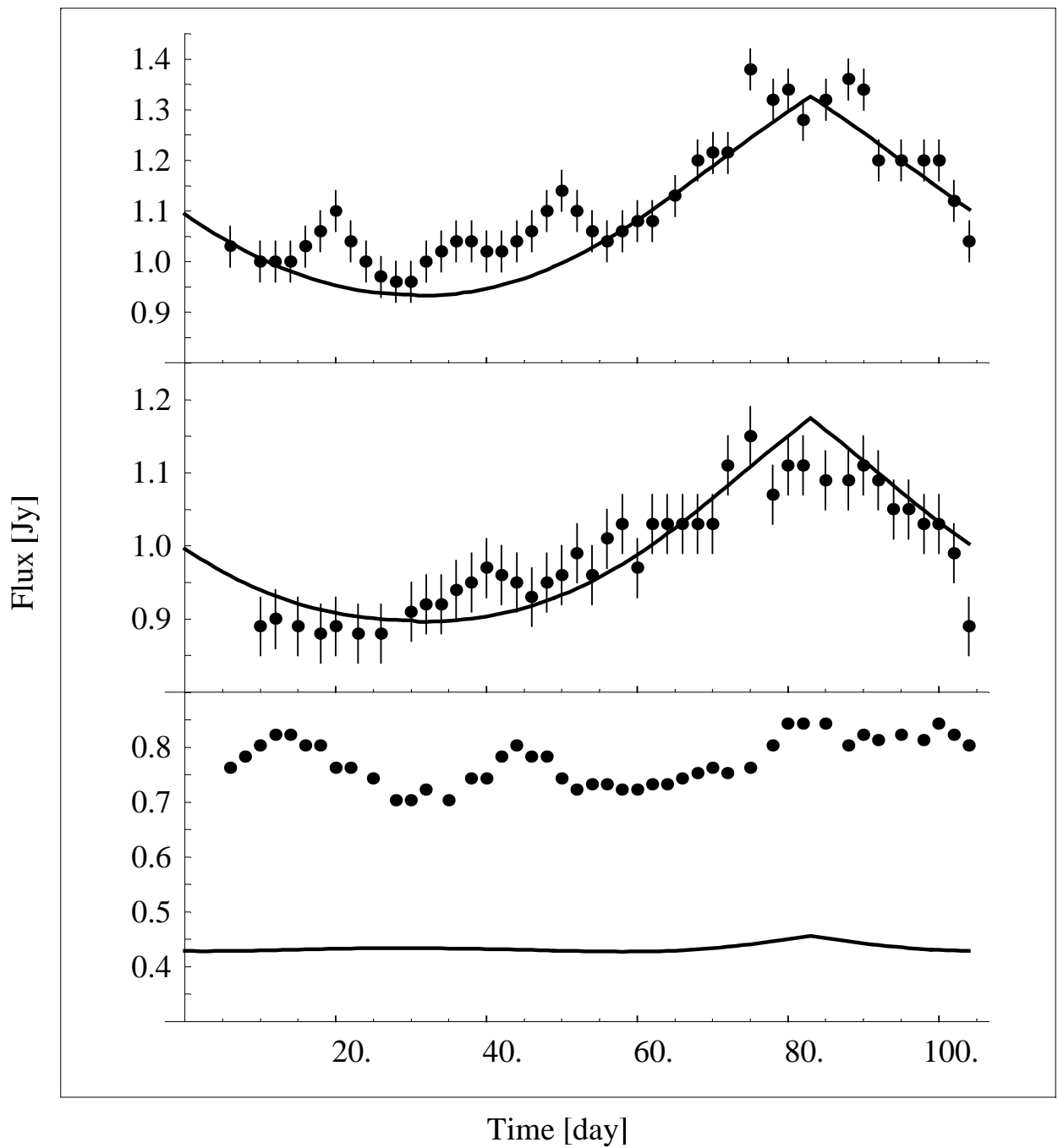}
\caption{Top panel: calculated light curve (solid) at $1.3$ cm, assuming a uniform nonthermal
halo with radius $R_h=21.5\,r_S$, enclosing a compact disk with radius $R_d=17.5\,r_S$. See text
for additional parameter values. The data are folded over a $106$-day period, and are taken from
Zhao et al. (2001). Middle panel: same as above, except here at a wavelength of $2.0$ cm.
Bottom panel: same as above, except here at a wavelength of $3.6$ cm.}
\label{fig:3}
\end{figure}

\newpage
\begin{figure}[ht]
\plotone{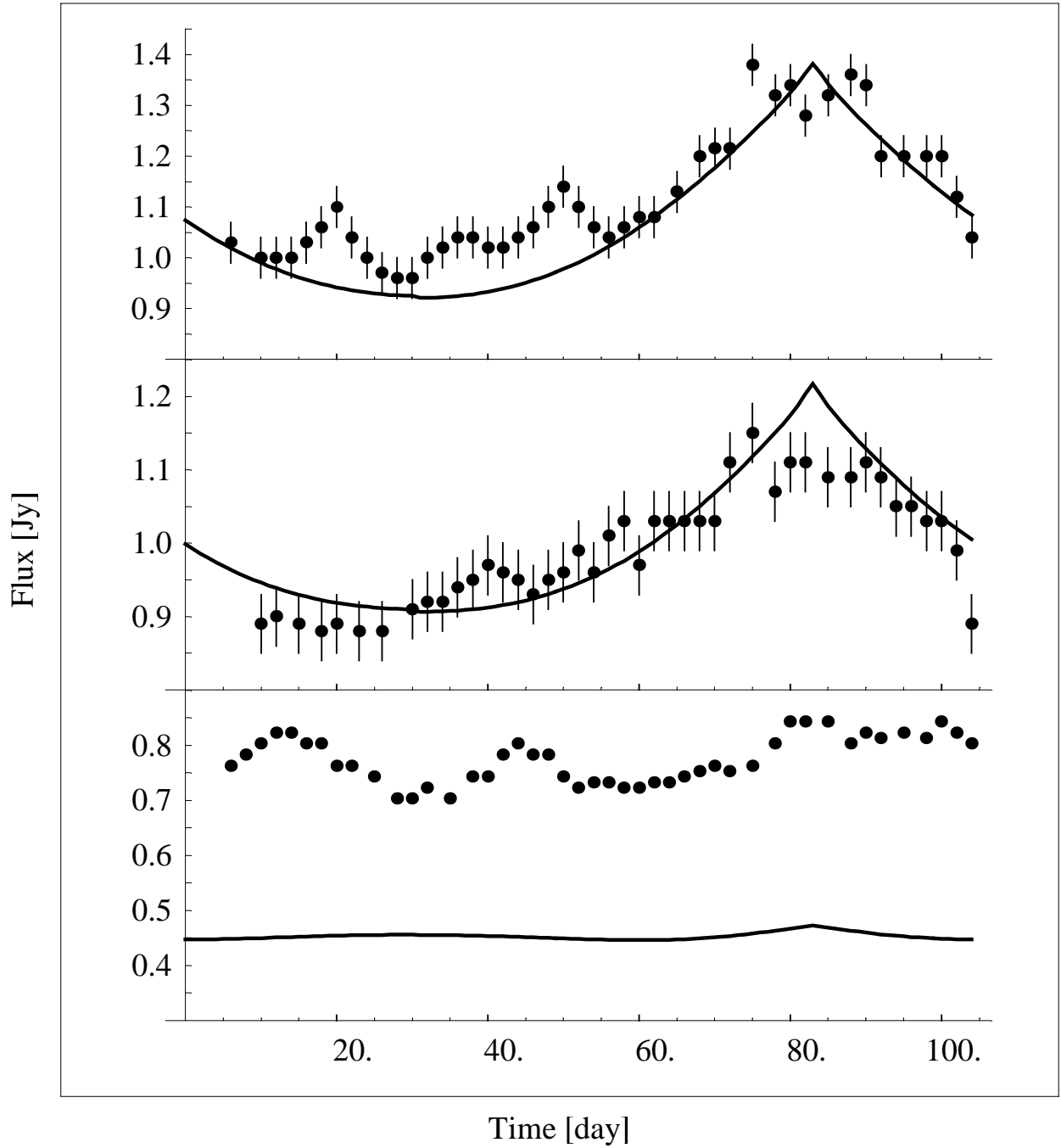}
\caption{Same as Figure 3, except here for a halo with a $1/r$ density
profile.  See text for additional parameter values.}
\label{fig:4}
\end{figure}

\newpage
\begin{figure}[ht]
\plotone{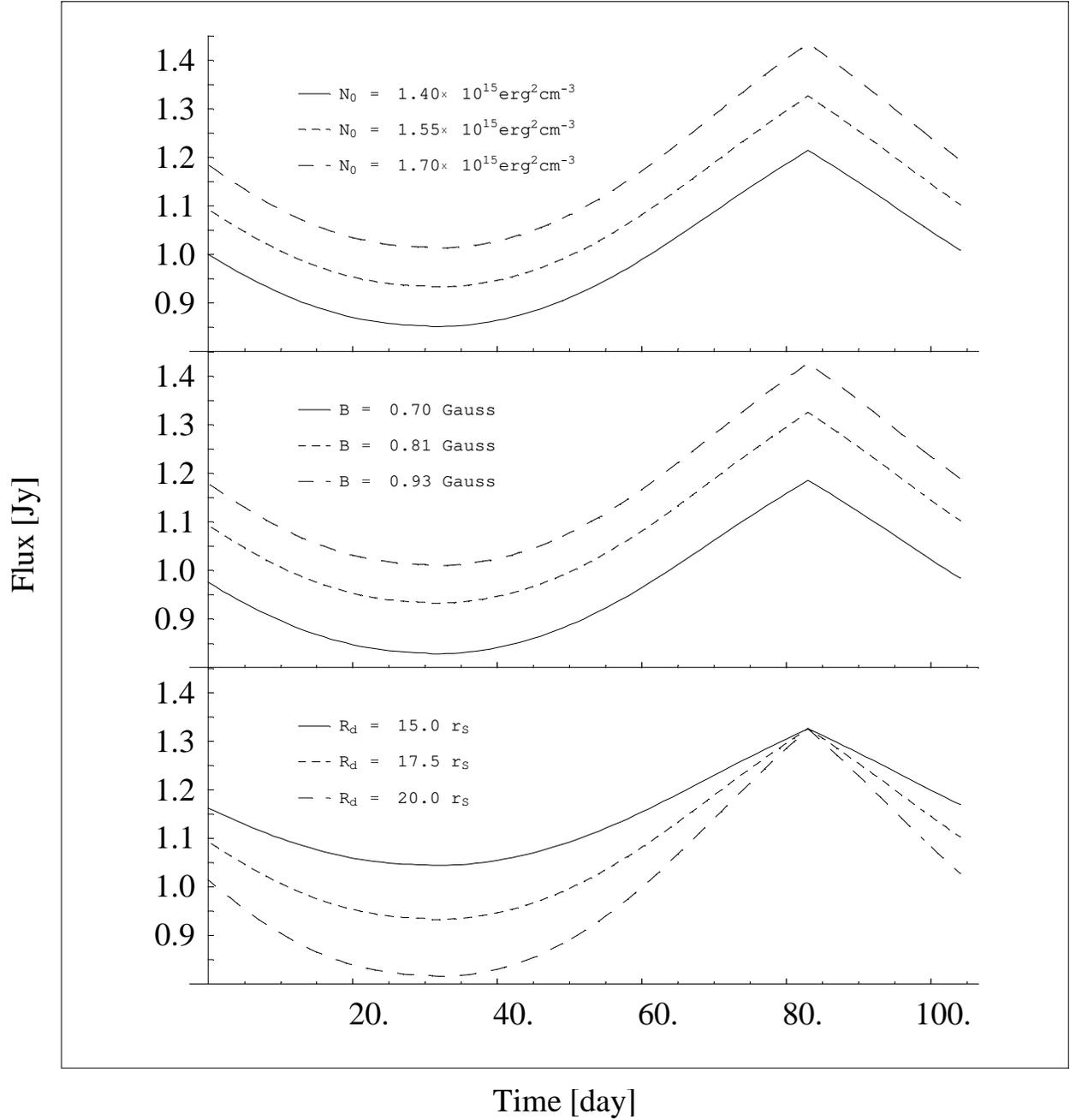}
\caption{Top panel: the effect on Sagittarius A*'s light curve at $1.3$ cm due
to variations (as shown) in the halo's nonthermal particle density.
Middle panel: the effect on Sagittarius A*'s light curve at $1.3$ cm due
to variations (as shown) in the halo's magnetic field.
Bottom panel: the effect on Sagittarius A*'s light curve at $1.3$ cm due
to variations (as shown) in the disk size.}
\label{fig:5}
\end{figure}



\newpage
\begin{thebibliography}{}

\bibliographystyle{apj}

\bibitem[Baganoff et al 2001]{Baganoff01}{Baganoff, F. et al. 2001,
Nature, 413, 45}

\bibitem[{Bower and Backer} 1998]{Bower98}
{Bower}, G.~C. and Backer, D.~C. 1998, \apjl, 496, L97

\bibitem[Bower et al. 1999]{Bower99}{Bower, G., Backer, D., Zhao, J., Goss,
M., \& Falcke, H. 1999, ApJ, 521, 582}

\bibitem[Bower et al. 2004]{Bower04}{Bower, G. C., Falcke, H., Hernstein, R., M.,
Zhao, J.-H., Goss, W. M., Backer, D. C. 2004, Science, 304, 704}

\bibitem[Bower et al 2001]{Bower01}{Bower, G., Wright, M., Falcke, H., \&
Backer, D. 2001, ApJ, 555, 103}

\bibitem[Bromley et al. 2001]{Bromley01}{Bromley, B., Melia, F., \& Liu, S.
2001, ApJ, 555, L83}

\bibitem[Coker et al 1997]{Coker97}{Coker, R.F., \& Melia, F. 1997, ApJ, 488, L49}

\bibitem[{Genzel et al.} 2003]{Genzel03}
{Genzel}, R., Sch\"odel, R., Ott, T. et al. 2003, \nat, 425, 934

\bibitem[{Kato} 1990]{Kato90} {Kato}, S. 1990, PASP, 42, 99

\bibitem[{Krichbaum et al.} 1999]{Krichbaum99}
{Krichbaum}, T.~P., Witzel, A., and Zensus, J.~A. 1999, in The Central Parsecs of
the Galaxy: ASP Conference Series 186, 89

\bibitem[{Liu and Melia} 2001]{Liu01} {Liu}, S. and Melia, F. 2001, \apjl, 561, L77

\bibitem[{Liu and Melia} 2002]{Liu02} {Liu}, S. and Melia, F. 2002, \apjl, 573, L23

\bibitem[Melia 1992]{Melia92_a}{Melia, F. 1992, ApJ, 387, L25}

\bibitem[Melia 1994]{Melia94}{Melia, F. 1994, ApJ, 577, 426}

\bibitem[Melia et al 2001]{Melia01b}{Melia, F., Bromley, B., \& Liu, S.
2001, ApJ, 554, L37}

\bibitem[{Melia and Falcke} 2001]{Melia01} {Melia}, F. and Falcke, H. 2001, ARAA, 39, 309

\bibitem[{Melia et al.} 1992]{Melia92_b}
{Melia}, F., Jokipii, J.~R., and Narayanan, A. 1992, \apjl, 395, L87

\bibitem[Melia et al. 2000]{Melia00}{Melia, F., Liu, S., \& Coker, R. 2000,
ApJ, 545, L117}

\bibitem[Nelson \& Papaloizou 2000]{Nelson00}{Nelson, R. P. \& Papaloizou, C. B. 2000,
MNRAS, 315, 570}

\bibitem[{Pacholczyk} 1970]{Pacholczyk70}
{Pacholczyk}, A.~G. 1970, Radio Astrophysics, Freeman: San Francisco

\bibitem[{Rockefeller et al.} 2005]{Rockefeller05}
{Rockefeller}, G., Fryer, C., and Melia, F. 2005, \apjl, submitted

\bibitem[{Rybicki and Lightman} 1979]{Rybicki79}
{Rybicki}, G.~B. and Lightman, A.~P. 1979, Radiative Processes in Astrophysics, Wiley: New York

\bibitem[{Zhao et al.} 2001]{Zhao01}
{Zhao}, J.~H., Bower, G.~C., and Goss, W. M. 2001, \apj, 547, L29

\end{thebibliography}
\end{document}